\PassOptionsToPackage{dvipsnames}{xcolor}
\documentclass{bmvc2k}

\title{Audio-Visual Separation with Hierarchical Fusion and Representation Alignment}
\runninghead{Hu, Lin, Huang, Hou, Chang, Jiao}{Audio-Visual Separation}

\addauthor{Han Hu$^{*}$}{hxh347@student.bham.ac.uk}{1}
\addauthor{Dongheng Lin$^{*}$}{d.lin.2@bham.ac.uk}{1}
\addauthor{Qiming Huang}{qxh366@student.bham.ac.uk}{1}
\addauthor{Yuqi Hou}{yxh029@student.bham.ac.uk}{1}
\addauthor{Hyung Jin Chang}{h.j.chang@bham.ac.uk}{1} 
\addauthor{Jianbo Jiao}{j.jiao@bham.ac.uk}{1}

\addinstitution{
The MIx Group,
School of Computer Science,
University of Birmingham\\
Birmingham, UK
}

\usepackage{multirow}
\usepackage{enumitem}
\usepackage{float}
\usepackage{booktabs}
\usepackage{comment}
\usepackage{amssymb}
\usepackage{pifont}
\usepackage{epigraph}
\usepackage{wrapfig}
\usepackage[section]{placeins} 
\usepackage{hyperref} 
\hypersetup{
  colorlinks=true,
  linkcolor=MidnightBlue,
  citecolor=OliveGreen,
  urlcolor=BrickRed
}

\def\eg{\emph{e.g}\bmvaOneDot}
\def\Eg{\emph{E.g}\bmvaOneDot}
\def\etal{\emph{et al}\bmvaOneDot}
\newcommand{\rev}[1]{\textcolor{black}{#1}}

\begin{document}
\maketitle
\def\thefootnote{*}\footnotetext{Equal contribution.} 

\begin{abstract}
Self-supervised audio-visual source separation leverages natural correlations between audio and vision modalities to separate mixed audio signals. 
In this work, we first systematically analyse the performance of existing multimodal fusion methods for audio-visual separation task, demonstrating that the performance of different fusion strategies is closely linked to the characteristics of the sound—middle fusion is better suited for handling short, transient sounds, while late fusion is more effective for capturing sustained and harmonically rich sounds.
We thus propose a hierarchical fusion strategy that effectively integrates both fusion stages.
In addition, training can be made easier by incorporating high-quality external audio representations, rather than relying solely on the audio branch to learn them independently. 
To explore this, we propose a representation alignment approach that aligns the latent features of the audio encoder with embeddings extracted from pre-trained audio models.
Extensive experiments on MUSIC, MUSIC-21 and VGGSound datasets demonstrate that our approach achieves state-of-the-art results, surpassing existing methods under the self-supervised setting.
We further analyse the impact of representation alignment on audio features, showing that it reduces modality gap between the audio and visual modalities. 
\rev{The project page is at: \url{ https://happy-new-bears.github.io/hfra-audiosep/}.}

\end{abstract}

\epigraph{\makebox[0.9\textwidth][c]{\hspace{-9mm}``Alone we can do so little; together we can do so much.''}}{\textit{— Helen Keller}}

\section{Introduction}
\label{sec:intro}

\begin{figure}
\centering
\begin{tabular}{c}
\bmvaHangBox{%
    \includegraphics[width=0.8\linewidth]{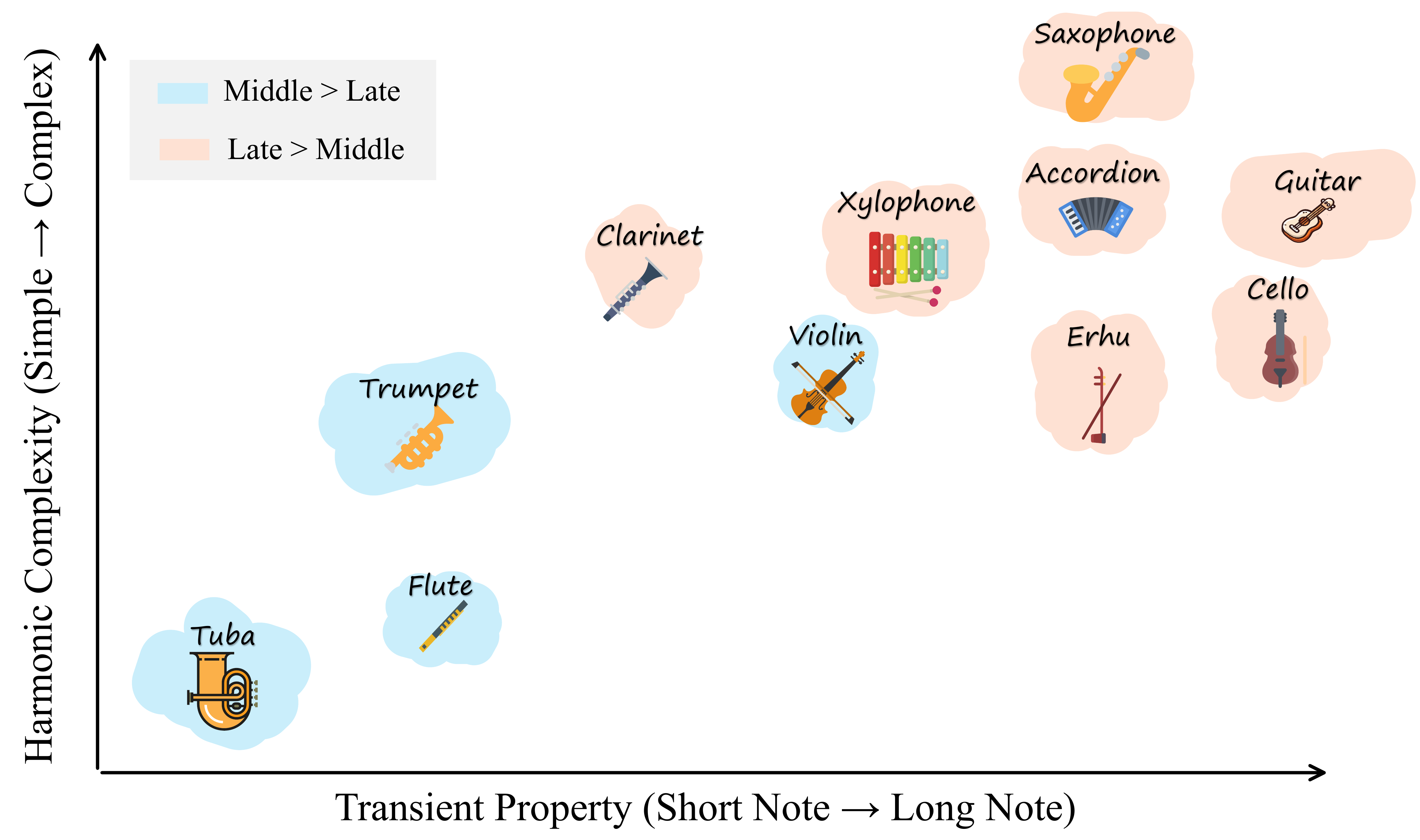}%
}\\
\end{tabular}
\caption{\textbf{Relationship Between Acoustic Properties of Musical Instruments and Fusion Strategies.}
Instruments with shorter transient properties and simpler harmonic structures are more suited to middle fusion. Conversely, instruments with sustained notes and complex harmonic structures benefit more from late fusion. Details can be found in Appendix \ref{app:s1}.} 
\label{fig:instrument_fusion}
\end{figure}

In our daily life, sounds come in diverse forms—some are quick and sharp, like a bird chirping or a raindrop hitting the ground, while others are smooth and lingering, like the deep hum of a cello or the steady strumming of a guitar. Transient sounds often carry distinct temporal signatures, while sustained sounds exhibit intricate harmonic structures that evolve over time \cite{Brattico2017GlobalSQ}. These differences in sound are not just something we hear; they also affect how we recognise and separate them in a mixed setting. 

Traditional audio separation methods rely solely on audio cues, struggling in complex environments where multiple sounds overlap \cite{molla2007single, nath2024separation}. 
In recent years, to improve the performance of audio separation, researchers have turned to visual modality as a strong prior \cite{Zhao_2018_ECCV}, known as ``audio-visual separation'', leveraging the natural correspondences between audio and visual signals.
As the visual modality is introduced, a natural question arises: How should the visual information be effectively integrated with the existing audio modality to enhance separation performance?  To address this, researchers have explored different fusion strategies to combine the two modalities.
Middle fusion integrates visual features at the bottleneck of the audio U-Net \cite{gao2019co}, while late fusion applies visual features at the final layer of the audio U-Net \cite{Zhao_2018_ECCV, dong2023clipsep}.

Figure \ref{fig:instrument_fusion} illustrates the relationship between \textbf{acoustic characteristics} and \textbf{fusion stages}.
Middle fusion performs better at capturing sharp, short-duration sounds but struggles with harmonically rich sources, while late fusion is more effective for continuous sounds but may overlook transient details. This trade-off naturally leads us to ask: Can we design a fusion strategy that combines the strengths of both approaches to improve separation \textit{across a wide range of sound characteristics?}
Our approach builds upon the idea that different sound characteristics require different levels of fusion. To this end, we propose a hierarchical fusion strategy that integrates both middle and late fusion. In this way, both short, sharp sounds and long, continuous ones are handled at the most suitable stages. 

In audio-visual separation, researchers have successfully leveraged large pre-trained vision models, such as CLIP \cite{radford2021learningtransferablevisualmodels}, to extract strong visual representations that significantly improve performance \cite{dong2023clipsep, wu2022wav2clip}. This naturally leads to the question: \textit{Can we apply large pre-trained audio models to benefit the audio separation task?} At first glance, it seems intuitive to replace the audio U-Net’s learned features with embeddings from a large pre-trained audio model. 
However, audio separation requires fine-grained time-frequency details to disentangle and reconstruct overlapping sounds, which the high-level audio embeddings from large pretrained models often lack. Thus, simply substituting learned features with pre-trained embeddings may not be a good choice.
Instead of directly replacing the audio latent representations at the bottleneck of U-Net, we propose representation alignment—an approach that aligns the latent space of the separation model with the embeddings from a large pre-trained audio model. In this way, the model can not only preserve fine-grained spectral details but also distil high-level semantic knowledge from the pre-trained embeddings.

To evaluate and understand the representation alignment approach, we investigate two key questions: whether aligning the U-Net’s latent features with a pre-trained model improves audio-visual separation, and if so, what underlying factors contribute to this improvement? Our experiment findings show that the proposed representation alignment method not only improves audio separation performance, but also enhances the semantic richness of audio latent features. Interestingly, representation alignment also reduces the modality gap between audio and visual representations, even though no explicit objective was introduced to enforce multimodal alignment.

We highlight the main contributions of this paper below:
\begin{itemize}[noitemsep, topsep=0pt]
    \item To our knowledge, this is the first study to reveal the correlation between acoustic characteristics of audio and different fusion strategies in audio-visual separation.
    \item We propose a self-supervised hierarchical fusion strategy for audio-visual separation.
    \item We propose a representation alignment loss that bridges the semantic gap between U-Net bottleneck features and pre-trained audio embeddings.
    \item Extensive experimental analysis validates the effectiveness of our approach, with performance gains on various benchmark datasets.
\end{itemize}

\section{Related Works}
\label{sec:related work}

\subsection{Audio  Visual Separation}
Existing audio-visual separation methods can be broadly categorized into self-supervised and weakly-supervised approaches. Self-supervised methods \cite{dong2023clipsep, zhao2019sound,Zhao_2018_ECCV} leverage the popular mix-and-separate strategy that creates synthetic audio mixtures by combining audio from different videos, enabling models to learn separation without the need for extra human annotations.
Weakly-supervised methods \cite{chen2023iquery,gao2019co} introduce additional semantic information, such as audio category labels, to provide indirect supervision. Compared to self-supervised methods, weakly-supervised methods can offer performance advantages but rely on additional annotations on audio categories, which increase labelling costs and may limit generalization to unseen scenarios. Thus, in this work, we focus on exploring self-supervised approaches. 

\rev{
While our method, like other self-supervised approaches, relies on global visual features, another line of work uses spatially grounded features via object detectors, \eg Co-Separation \cite{gao2019co}, CCoL \cite{tian2021cyclic}, and iQuery \cite{chen2023iquery}.
However, these methods have limitations that distinguish them from the self-supervised, global-feature-based approach: their performance is heavily dependent on the detector's accuracy, and they are restricted to categories the detector is trained on. For example, CCoL's masks can be blurry, while iQuery had to use a more general detector to accommodate new instruments in the MUSIC-21 dataset.
In contrast, our approach is purely self-supervised, leveraging global video-level features and natural audio-visual correspondences for separation, which makes it more robust to object detection failures and removes the dependency on object-level annotations or detectors.
}

Another important design choice in audio-visual separation models is the stage at which audio and visual features are fused. Existing methods primarily follow two fusion strategies: late fusion and middle fusion. In late fusion \cite{dong2023clipsep, zhao2019sound,Zhao_2018_ECCV}, visual features are applied at the final stages of the U-Net decoder to reweight the audio spectrogram. Middle fusion integrates visual embeddings earlier in the network by tiling and concatenating them with the bottleneck features of the U-Net \cite{ronneberger2015u}.
Prior studies suggest that late fusion generally outperforms middle fusion in self-supervised settings, while middle fusion has been shown to be beneficial in weakly-supervised scenarios \cite{tian2021cyclic, gao2019co}, particularly when combined with additional supervision signals, such as classification losses derived from labelled data \cite{chatterjee2022learning, chatterjee2021visual}. Different from previous works, we investigate how different sound characteristics influence the effectiveness of middle and late fusion and propose a hierarchical fusion strategy that combines both fusion mechanisms.

\subsection{Cross-Modal Representation Learning}
The CLIP model~\cite{radford2021learningtransferablevisualmodels}, based on contrastive learning, has been widely used as a pretraining framework to build joint embedding spaces for text and image modalities. Its success has also inspired extensions to the audio domain.
For instance, \cite{wu2022wav2clip} proposed a self-supervised approach where an additional audio encoder is trained to align input audio with the pretrained CLIP embedding space, enabling audio representations to inherit the multimodal alignment capabilities of CLIP. 
Similarly, in \cite{dong2023clipsep}, the authors explore the zero-shot modality transfer capability of CLIP by keeping the pre-trained model frozen while optimising only the remaining components for the target sound separation task. In our work, we follow the approach of \cite{dong2023clipsep}, which uses CLIP to extract visual features.

Similarly, the CLAP model \cite{laionclap2023, htsatke2022} extends such contrastive pretraining paradigm to audio-text embedding space
CLAP has been applied to audio-visual segmentation and detection tasks \cite{guo2024open, nakada2024deteclap}. However, its potential for audio-visual separation remains underexplored.

\section{Method}
\label{sec:method}

\begin{figure*}
\centering
\begin{tabular}{c}
\bmvaHangBox{%
    \includegraphics[width=0.95\linewidth]{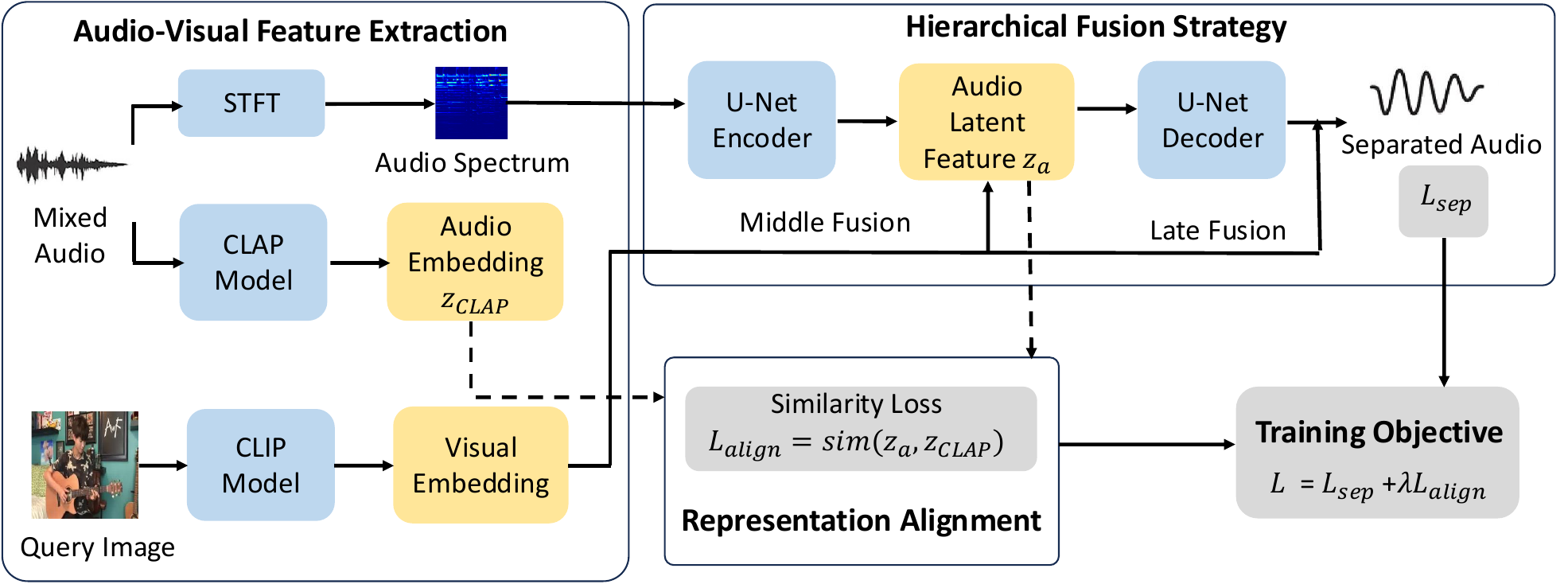}%
}\\
\end{tabular}
\caption{\textbf{Pipeline of our proposed method.} The pipeline consists of three key components: audio-visual feature extraction, hierarchical fusion, and representation alignment. It takes an audio mixture and corresponding video frames as input. The Audio-Visual Feature Extraction module processes the input through dedicated encoders, extracting audio features from spectrograms and CLAP and extracting visual features from CLIP. Hierarchical fusion includes middle fusion and late fusion, which happens at the bottleneck of the audio U-Net and the final layer of the audio decoder separately.} 
\label{fig:frame_work}
\end{figure*}

\subsection{Problem Definition}
\label{sec:problem_definition}
Audio separation aims to isolate individual sound sources from a mixed audio signal. During training, the model takes as inputs an audio mixture $\mathbf{x} = \sum_{i=1}^{n} s_i$, where $s_1, \dots, s_n$ are the $n$ audio tracks, along with their corresponding images $\mathbf{y}_1, \dots, \mathbf{y}_n$ extracted from the videos. 
We first transform the audio mixture $\mathbf{x}$ into a magnitude spectrogram $\mathbf{X} = |\text{STFT}(\mathbf{x})|$ and pass the spectrogram through an audio U-Net~\cite{ronneberger2015u}  to produce $k$ ($\geq n$) intermediate masks $\tilde{M}_1, \dots, \tilde{M}_k$.
On the other stream, each image $\mathbf{y}_i$ is encoded into an embedding $\mathbf{e}_i = \text{Enc}_{img} (\mathbf{y}_i) $. 

Middle fusion and late fusion integrate visual information at different stages of the separation model.
The process of middle fusion can be written as follows:
\begin{equation}
\tilde{M}_i = \text{Dec}_{a} \left( \text{Tile}(\mathbf{e}_i) \oplus \text{Enc}_{a}(\mathbf{X}) \right).
\end{equation}

The process of late fusion can be written as follows:
\begin{equation}
\tilde{M}_i = \text{Proj}_{1}(\mathbf{e}_i) \odot \text{Dec}_{a}(\text{Enc}_{a}(\mathbf{X})),
\end{equation}
where the function $\text{Enc}_{a}(\cdot)$ denotes the audio encoder, which extracts bottleneck audio features, while $\text{Dec}_{a}(\cdot)$ represents the audio decoder.
In middle fusion, $\text{Tile}(\mathbf{e}_i)$ expands the visual embedding spatially to match the dimensions of $\text{Enc}_{a}(\mathbf{X})$, and $\oplus$ denotes the channel-wise concatenation of visual and audio features. Late fusion applies a projection function $\text{Proj}_{1}(\mathbf{e}_i)$ mapping the visual embedding to a compatible space before performing element-wise multiplication ($\odot$) with the decoded audio representation.

The separated spectrogram \(\hat{\mathbf{X}}_i\) for each source \(i\) is then obtained by multiplying the estimated mask \(\tilde{M}_i\) to the  input mixed spectrogram \(\mathbf{X}\) through element-wise multiplication:
\begin{equation}
\hat{\mathbf{X}}_i = \tilde{M}_i \odot \mathbf{X}.
\end{equation}

The separation loss \(\mathcal{L}_{\text{sep}}\) is then defined using the \(L_1\) distance between the predicted spectrogram \(\hat{\mathbf{X}}_i\) and the ground-truth spectrogram \(\mathbf{X}_i\):
\begin{equation}
\mathcal{L}_{\text{sep}} = \sum_{i=1}^{n} \|\hat{\mathbf{X}}_i - \mathbf{X}_i\|_1.
\end{equation}

\subsection{Hierarchical fusion for Audio-Visual Separation}
\label{sec:fusion}
We can observe from Figure~\ref{fig:instrument_fusion} that middle and late fusion strategies exhibit complementary advantages depending on the characteristics of the target sound. In 
Specifically, middle fusion is more effective at separating transient sound instruments (\eg trumpet, flute) and stable low-frequency instruments (\eg tuba), whereas late fusion performs better for sustained instruments (\eg saxophone, cello) and harmonically rich instruments (\eg acoustic guitar, xylophone). 
Appendix \ref{sec:s3} further provides a quantitative analysis of the relationship between separation performance and fusion stage.
Motivated by such observation, we propose a hierarchical fusion strategy that integrates both middle and late fusion to leverage their complementary strengths.
The hierarchical fusion mechanism is formulated as:
\begin{equation}
\tilde{M}_i = \text{Proj}_{1}(\mathbf{e}_i) \odot \text{Dec}_{a} \left( \text{Tile}(\text{Proj}_{2}(\mathbf{e}_i) )\oplus \text{Enc}_{a}(\mathbf{X}) \right),
\end{equation}
where \(\text{Proj}_{1}\) and \(\text{Proj}_{2}\) are both single fully connected layers that project the visual embedding \(\mathbf{e}_i\) into appropriate latent spaces for different fusion mechanisms.

\subsection{Audio Representation Alignment}
\label{sec:audio_representation_alignment}

Previous studies have shown that using CLIP-extracted visual features greatly improves audio-visual separation \cite{dong2023clipsep} because CLIP learns strong multimodal representations that align well with semantic categories. Inspired by this, we hypothesise that learning high-quality audio representations can similarly enhance source generation performance in audio-visual separation. In this way, we propose audio representation alignment, which encourages the U-Net encoder’s latent representations to align with self-supervised audio embeddings extracted by the pretrained large audio model.

The audio representation alignment loss function is defined as follows:

\begin{equation}
\mathcal{L}_{\text{audio\_align}} = 1 - \text{sim}(\mathbf{z}_a, \mathbf{z}_{\text{CLAP}}),
\end{equation}
where the latent representations are defined as \( \mathbf{z}_a = \text{Enc}_a(\mathbf{X}) \) and \( \mathbf{z}_{\text{CLAP}} = \text{Enc}_{\text{CLAP}}(\mathbf{x}) \). 

Here, \( \text{Enc}_a(\cdot) \) denotes the audio encoder of the U-Net, which extracts latent features from the input spectrogram \( \mathbf{X} \), and \( \text{Enc}_{\text{CLAP}}(\cdot) \) represents the pretrained self-supervised audio encoder from CLAP, which generates a semantic embedding from the waveform of the same mixed audio input \( \mathbf{x} \). The function \( \text{sim}(\cdot, \cdot) \) is thecosine similarity function:
\begin{equation}
\text{sim}(\mathbf{z}_a, \mathbf{z}_{\text{CLAP}}) = \frac{\mathbf{z}_a \cdot \mathbf{z}_{\text{CLAP}}}{\|\mathbf{z}_a\| \|\mathbf{z}_{\text{CLAP}}\|}.
\end{equation}

In practice, we add this term to the original audio separation objectives described in Section~\ref{sec:problem_definition}. Therefore, the overall training loss is:
\begin{equation}
\mathcal{L} = \mathcal{L}_{\text{sep}} + \lambda \mathcal{L}_{\text{audio\_align}},
\end{equation}
where \( \lambda > 0 \) is a hyperparameter that controls the tradeoff between audio separation performance and representation alignment. 


\section{Experiments}
In this section, we conducted experiments not only to demonstrate that our proposed method can improve the performance of audio-visual separation but also to gain deeper insights into \textbf{why} these improvements occur.

\subsection{Experimental Settings}
\paragraph{Datasets.} We conduct experiments on three widely-used datasets: MUSIC \cite{Zhao_2018_ECCV}, MUSIC-21 \cite{zhao2019sound} and VGGSound~\cite{chen2020vggsound}. The MUSIC dataset contains  601 untrimmed videos of musical solos and duets across 11 categories of musical instruments, due to some videos becoming unavailable online over time.
Following the data splits of \cite{Zhao_2018_ECCV}, we use 483 videos for training and 118 videos for testing, with the test set exclusively consisting of solo performances.
The MUSIC-21 dataset \cite{zhao2019sound} consists of solo videos across 21 instrument categories. We utilise 1,039 online available solo videos and adopt the training/testing split of \cite{zhao2019sound}, with 831 videos for training and 208 for testing.
VGGSound~\cite{chen2020vggsound} is a large-scale audio-visual dataset consisting of 10-second video clips ``in the wild'', covering 309 sound event categories. Our training set contains 132{,}760 videos, and our test set includes 11{,}147 videos.

\paragraph{Baselines.}
We compare our method with several recent self-supervised approaches across all three datasets. On the MUSIC dataset, we include: NMF-MFCC \cite{virtanen2007monaural}, a non-learnable audio-only baseline (results from \cite{gao2019co}); Sound-of-Pixels \cite{Zhao_2018_ECCV}; CLIPSep \cite{dong2023clipsep}, retrained under the same settings; and Semantic Grouping Network (SGN) \cite{mo2024semanticgroupingnetworkaudio}. For MUSIC-21, we compare against NMF-MFCC, Sound-of-Pixels, and CLIPSep following the same protocol. On VGGSound, we evaluate our method against CLIPSep as the main baseline, given its strong performance in audio-visual separation.

\paragraph{Evaluation Metrics.} We assess the performance of sound separation using three standard metrics: Signal-to-Distortion Ratio (SDR), Signal-to-Interference Ratio (SIR), and Signal-to-Artefacts Ratio (SAR). SIR evaluates how well interfering sources are suppressed, while SAR reflects the level of artefacts introduced during separation \cite{metrics06}. Among these, SDR is generally considered the most important metric, as it provides an overall measure of separation quality \cite{petermann2020deeplearningbasedsource} by accounting for both interference and artefacts. 
In summary, higher values for all three metrics indicate better performance, while we primarily focus on the SDR.

\paragraph{Implementation Details.} 

\begin{table}
\centering
\caption{\textbf{Audio-visual separation performance comparison on the MUSIC dataset.} Best results in \textbf{bold}, second-best \underline{underlined}.\rev{$^{\dag}$Results taken from \cite{chen2023iquery}.}}
\label{tab:music_results}
\begin{tabular}{lccc}
\toprule
\textbf{Method} & \textbf{SDR}~$\uparrow$ & \textbf{SIR}~$\uparrow$ & \textbf{SAR}~$\uparrow$ \\
\midrule
RPCA \cite{huang2012singing} $^{\dag}$               & $-0.62$ & $2.32$ & $2.41$ \\
Wave-U-Net \cite{stoller2018wave}  $^{\dag}$         & $3.80$  & $6.75$ & $6.62$ \\
ResUNetDecouple+ \cite{kong2021decoupling} $^{\dag}$   & $3.98$  & $7.17$ & $6.91$ \\
MP-Net \cite{yu2019free} $^{\dag}$                    & $4.82$  & $10.19$ & $10.56$  \\
SGN \cite{mo2024semanticgroupingnetworkaudio} $^{\dag}$  & $5.20$  & $10.81$ & $10.67$ \\
\midrule
NMF-MFCC \cite{virtanen2007monaural}     & $0.92$  & $5.68$  & $6.84$ \\
Sound-of-Pixels \cite{Zhao_2018_ECCV}     & $3.84$  & $9.66$  & $9.32$ \\
CLIPSep \rev{(Middle Fusion)} \cite{dong2023clipsep} & $5.57$  & \textbf{12.99} & $8.60$ \\
CLIPSep \rev{(Late Fusion)} \cite{dong2023clipsep}   & \underline{5.86} & $11.65$ & \underline{9.72} \\
\midrule
\textbf{Ours} (Hierarchical + Align)      & \textbf{6.72} & \underline{12.60} & \textbf{10.21} \\
\bottomrule
\end{tabular}
\end{table}

All audio is resampled to 11,625 Hz. For MUSIC and MUSIC-21, we extract 5-second segments from each video (random crop for training, centre crop for testing). VGGSound clips are 10 seconds long and used without cropping to retain real-world diversity.
Audio is processed using STFT with a Hann window of size 1024 and hop length of 256, producing complex spectrograms. We then compute the magnitude and apply a logarithmic mapping along the frequency axis to reflect human perception. The final input is a $256 \times 256$ log-scaled magnitude spectrogram, where $T = 256$ time frames and $F = 256$ frequency bins.
Visual inputs are sampled at 8 frames per second. A single frame per clip is extracted (randomly for training, centre for testing) and encoded using a frozen CLIP ViT-B/32 to obtain a 512-dimensional embedding.
We train our model using a batch size of 32. The learning rate is initially set to $10^{-4}$ and reduced by a factor of 0.1 at the 60th epoch.

\subsection{Audio-Visual Sound Source Separation Results}

\paragraph{Quantitative Evaluation.}
We evaluate our method on MUSIC, MUSIC-21, and VGGSound, with results summarized in Table~\ref{tab:music_results}, Table~\ref{tab:music21_results}, and Table~\ref{tab:vgg_results1}, respectively. Overall, our approach consistently outperforms existing baselines, especially in terms of SDR, which is widely regarded as the most important metric for source separation. 

These results validate the effectiveness of our design in handling a wide range of audio-visual scenarios.

\begin{table}
\centering
\caption{\textbf{Audio-visual separation results on MUSIC21 dataset.} Best results in \textbf{bold}, second-best \underline{underlined}. $^{\dag}$Results taken from \cite{chen2023iquery}. $^{\ddag}$Results taken from \cite{Islam2024VGASS}.}
\label{tab:music21_results}
\begin{tabular}{lccc}
\toprule
\textbf{Methods} & \textbf{SDR}~$\uparrow$ & \textbf{SIR}~$\uparrow$ & \textbf{SAR}~$\uparrow$ \\
\midrule
NMF-MFCC \cite{virtanen2007monaural}$^{\dag}$     & 2.78 & 6.70 & 9.21 \\
AV-MMix-and-Separate \cite{gao2019co}$^{\ddag}$   & 3.23 & 7.01 & 9.14 \\
Sound-of-Pixels \cite{Zhao_2018_ECCV}             & 6.51 & 12.84 & 10.58 \\
CLIPSep \rev{(Middle Fusion)} \cite{dong2023clipsep}    & 7.36 & \textbf{14.28} & 10.22 \\
CLIPSep \rev{(Late Fusion)} \cite{dong2023clipsep}      & 7.27 & 13.10 & \underline{11.14} \\
\midrule
\textbf{Ours} (Hierarchical)                      & \underline{7.72} & 13.63 & 10.94 \\
\textbf{Ours} (Hierarchical + Align)              & \textbf{8.03} & \underline{13.92} & \textbf{11.36} \\
\bottomrule
\end{tabular}
\end{table}

\begin{table}
\centering
\caption{\textbf{Audio-visual separation results on VGGSound.} Best results in \textbf{bold}, second-best \underline{underlined}.}
\label{tab:vgg_results1}
\begin{tabular}{lccc}
\toprule
\textbf{Method} & \textbf{SDR}~$\uparrow$ & \textbf{SIR}~$\uparrow$ & \textbf{SAR}~$\uparrow$ \\
\midrule
CLIPSep \rev{(Late Fusion)}         & 0.90 & 7.47 & \underline{8.26} \\
CLIPSep \rev{(Middle Fusion)}     & \underline{1.16} & \textbf{9.41} & 6.95 \\
\textbf{Ours} (Hierarchical + Align) & \textbf{1.97} & \underline{8.96} & \textbf{9.62} \\
\bottomrule
\end{tabular}
\end{table}

\rev{While our method achieves state-of-the-art SDR scores, we observe a slightly lower SIR compared to some baselines, such as CLIPSep with Middle Fusion on the MUSIC-21 and VGGSound datasets. We attribute this to the trade-off between suppressing interfering sources (SIR) and avoiding the introduction of artefacts (SAR). Our approach prioritises reducing artefacts, as evidenced by our consistently high SAR scores across all datasets, particularly on MUSIC and VGGSound. Since SDR provides a comprehensive measure of separation quality by accounting for both interference and artefacts, our superior SDR performance indicates that the gain from reducing artefacts outweighs the minor decrease in interference suppression, leading to a higher overall separation quality.}

\paragraph{Qualitative Evaluation.}

\begin{figure}
\centering
\begin{tabular}{c}
\bmvaHangBox{%
    \includegraphics[width=0.8\linewidth]{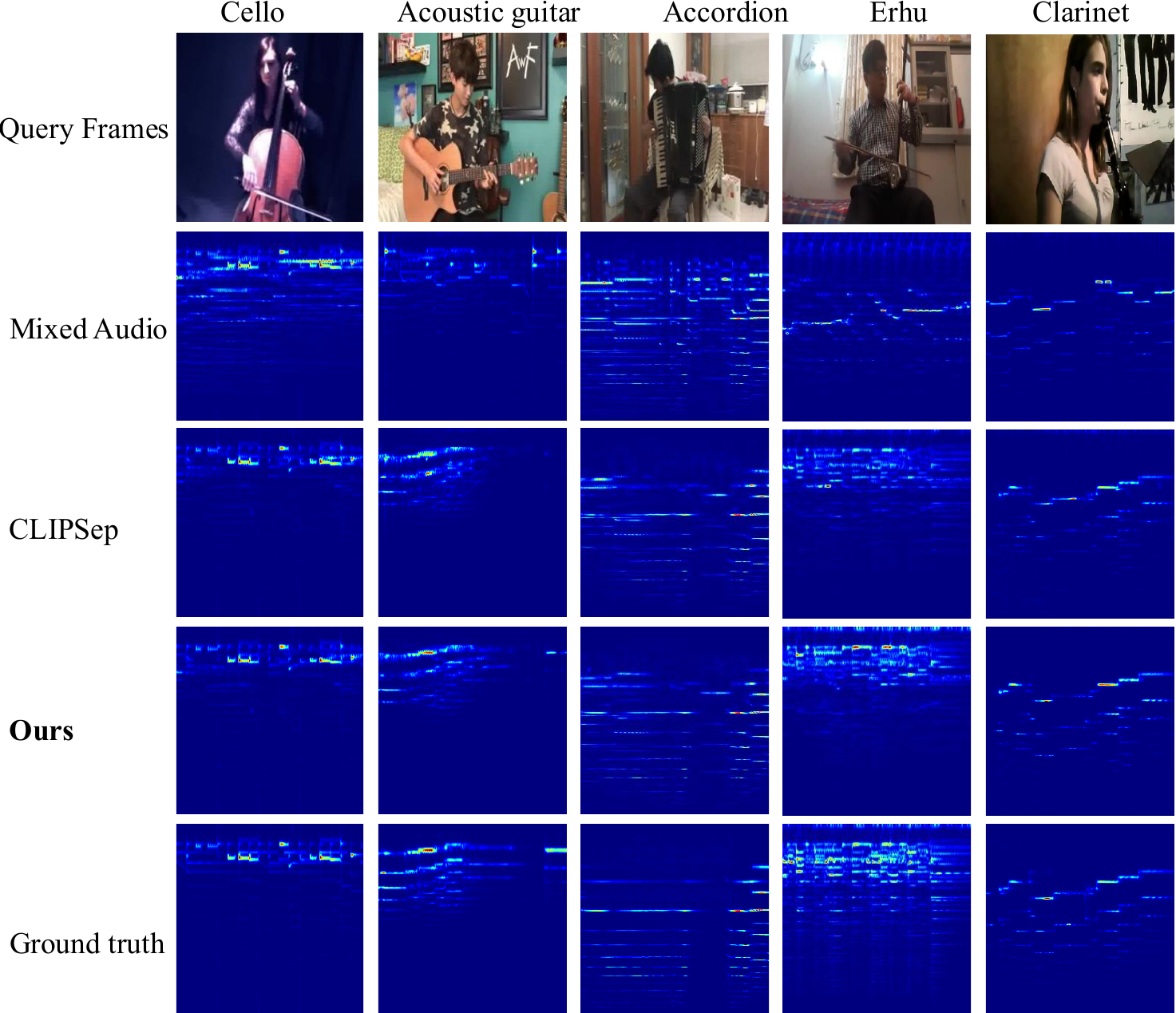}%
}\\
\end{tabular}
\caption{\textbf{Qualitative Performance on MUSIC dataset.}  We compared our method (the fourth row) with \citet{dong2023clipsep} (the third row). More results in Appendix \ref{app:more_examples}.}

\FloatBarrier

\label{fig:qualitative_music}
\end{figure}

Figure~\ref{fig:qualitative_music} provides qualitative examples of sound separation on the MUSIC dataset.  Our method produces cleaner and more distinct separated sounds, reducing interference. The visualisation results further support that our approach can separate mixed musical components with high precision.

\subsection{Audio Representation and Modality Gap Evaluation}

To evaluate the effect of representation alignment, we perform linear probing on the MUSIC test set. As shown in Table~\ref{tab:linear_probe}, for the U-Net latent feature, classification accuracy significantly improves after using representation alignment, indicating that aligned U-Net features capture more semantic information.

\begin{table}[!htbp]
\centering
\caption{\textbf{Linear probing accuracy on different audio representations.}}
\label{tab:linear_probe}
\begin{tabular}{llc}
\toprule
\textbf{Method} & \textbf{Representation Source} & \textbf{Accuracy (\%)} \\
\midrule
w/o Alignment   & U-Net Bottleneck   & 49.58 \\
w/ Alignment    & U-Net Bottleneck   & 58.82 \\
Frozen CLAP     & CLAP Embedding     & 59.66 \\
\bottomrule
\end{tabular}
\end{table}

Although classification accuracy remains slightly lower than that of frozen CLAP embeddings, this is reasonable. CLAP is trained explicitly for semantic discrimination, while the U-Net encoder is optimised for source separation, focusing more on preserving fine-grained spectral details than on maximising classification accuracy.

We further calculate the modality gap between audio and visual representations on the MUSIC testset following \cite{ModalityGap}. 
The results are shown in Table~\ref{tab:modality_gap}.
\rev{
Given audio embeddings \( \mathbf{X} = \{ x_i \}_{i=1}^{N} \) and visual embeddings \( \mathbf{Y} = \{ y_i \}_{i=1}^{N} \),  the modality gap is defined as follows: }

\begin{equation}
    \Delta_{\text{gap}} = \frac{1}{N} \sum_{i=1}^{N} x_i - \frac{1}{N} \sum_{i=1}^{N} y_i.
\end{equation}


\begin{wraptable}[9]{r}{0.46\textwidth}
\centering
\caption{\textbf{Comparison of modality gap with and without representation alignment.}}
\label{tab:modality_gap}
\begin{tabular}{lc}
\toprule
\textbf{Method} & \textbf{Modality Gap} \\
\midrule
w/o Alignment & 1.171 \\
w/ Alignment  & 0.976 \\
\bottomrule
\end{tabular}
\end{wraptable}

 We can observe that representation alignment leads to an obvious reduction in modality gap. Interestingly, this reduction occurs despite the absence of an explicit objective to minimise the gap between audio and visual modalities. Although the alignment process is not directly designed for this purpose, it implicitly promotes better interaction between audio and visual features, ultimately benefiting the source separation task.

\subsection{Ablation Study}

Table~\ref{tab:ablation_study} presents an ablation study evaluating the impact of visual backbone, fusion strategy, and representation alignment on audio-visual separation performance.
Results show that hierarchical fusion consistently yields higher SDR and SAR scores, while maintaining a competitive SIR score. Additionally, the combination of hierarchical fusion and alignment achieves the best overall results.

\begin{table}[!htbp]
\centering
\caption{\textbf{Ablations on fusion strategies and with or without representation alignment.}}
\label{tab:ablation_study}
\begin{tabular}{llcccc}
\toprule
\textbf{Visual Backbone} & \textbf{Fusion} & \textbf{Alignment} & \textbf{SDR}~$\uparrow$ & \textbf{SIR}~$\uparrow$ & \textbf{SAR}~$\uparrow$ \\
\midrule
ResNet-18 & Late          & \ding{55}  & 3.84 & 9.66  & 9.32 \\
CLIP ViT-B/32 & Middle    & \ding{55}  & 5.57 & \textbf{12.99} & 8.60 \\
CLIP ViT-B/32 & Late      & \ding{55}  & 5.86 & 11.64 & 9.72 \\
CLIP ViT-B/32 & Late      & \ding{51} & 6.07 & 12.10 & 9.73 \\
CLIP ViT-B/32 & Hierarchical & \ding{55}  & 6.65 & 12.39 & 10.11 \\
CLIP ViT-B/32 & Hierarchical & \ding{51} & \textbf{6.72} & 12.60 & \textbf{10.21} \\
\bottomrule
\end{tabular}
\end{table}

\vspace{-2.5em} 
\section{Conclusion}

In this work, we analyse the complementary strengths of middle and late fusion in audio-visual separation and propose a hierarchical fusion strategy that leverages both to improve performance across diverse scenarios. To further enhance semantic understanding, we introduce a representation alignment mechanism that aligns U-Net audio features with CLAP embeddings, enriching semantic information while preserving spectral details. Experiments confirm that both components contribute to separation performance. While our model uses global CLIP features without explicit sound source localisation, future work can explore self-supervised localisation to guide visual attention toward sounding objects and further improve separation performance.

\section*{Acknowledgements}

This project is partially supported by the Royal Society grants (SIF\textbackslash R1\textbackslash231009, IES\textbackslash R3\textbackslash223050) and an Amazon Research Award.
The computations in this research were performed using the Baskerville Tier 2 HPC service. Baskerville was funded by the EPSRC and UKRI through the World Class Labs scheme (EP\textbackslash T022221\textbackslash1) and the Digital Research Infrastructure programme (EP\textbackslash W032244\textbackslash1) and is operated by Advanced Research Computing at the University of Birmingham.


\bibliography{egbib}

\begin{thebibliography}{30}
\providecommand{\natexlab}[1]{#1}
\providecommand{\url}[1]{\texttt{#1}}
\expandafter\ifx\csname urlstyle\endcsname\relax
  \providecommand{\doi}[1]{doi: #1}\else
  \providecommand{\doi}{doi: \begingroup \urlstyle{rm}\Url}\fi

\bibitem[Brattico et~al.(2017)Brattico, Brattico, and Vuust]{Brattico2017GlobalSQ}
Pauli Brattico, Elvira Brattico, and Peter Vuust.
\newblock Global sensory qualities and aesthetic experience in music.
\newblock \emph{Frontiers in Neuroscience}, 11, 2017.
\newblock URL \url{https://api.semanticscholar.org/CorpusID:12298812}.

\bibitem[Caclin et~al.(2005)Caclin, McAdams, Smith, and Winsberg]{caclin2005acoustic}
Anne Caclin, Stephen McAdams, Bennett~K Smith, and Suzanne Winsberg.
\newblock Acoustic correlates of timbre space dimensions: A confirmatory study using synthetic tones.
\newblock \emph{The Journal of the Acoustical Society of America}, 118\penalty0 (1):\penalty0 471--482, 2005.

\bibitem[Chatterjee et~al.(2021)Chatterjee, Le~Roux, Ahuja, and Cherian]{chatterjee2021visual}
Moitreya Chatterjee, Jonathan Le~Roux, Narendra Ahuja, and Anoop Cherian.
\newblock Visual scene graphs for audio source separation.
\newblock In \emph{Proceedings of the IEEE/CVF International Conference on Computer Vision}, pages 1204--1213, 2021.

\bibitem[Chatterjee et~al.(2022)Chatterjee, Ahuja, and Cherian]{chatterjee2022learning}
Moitreya Chatterjee, Narendra Ahuja, and Anoop Cherian.
\newblock Learning audio-visual dynamics using scene graphs for audio source separation.
\newblock \emph{Advances in Neural Information Processing Systems}, 35:\penalty0 16975--16988, 2022.

\bibitem[Chen et~al.(2020)Chen, Xie, Vedaldi, and Zisserman]{chen2020vggsound}
H.~Chen, W.~Xie, A.~Vedaldi, and A.~Zisserman.
\newblock Vgg-sound: A large-scale audio-visual dataset.
\newblock In \emph{IEEE International Conference on Acoustics, Speech and Signal Processing (ICASSP)}, pages 721--725. IEEE, 2020.

\bibitem[Chen et~al.(2023)Chen, Zhang, Lian, Yang, Zeng, and Shi]{chen2023iquery}
Jiaben Chen, Renrui Zhang, Dongze Lian, Jiaqi Yang, Ziyao Zeng, and Jianbo Shi.
\newblock iquery: Instruments as queries for audio-visual sound separation.
\newblock In \emph{Proceedings of the IEEE/CVF Conference on Computer Vision and Pattern Recognition}, pages 14675--14686, 2023.

\bibitem[Chen et~al.(2022)Chen, Du, Zhu, Ma, Berg-Kirkpatrick, and Dubnov]{htsatke2022}
Ke~Chen, Xingjian Du, Bilei Zhu, Zejun Ma, Taylor Berg-Kirkpatrick, and Shlomo Dubnov.
\newblock Hts-at: A hierarchical token-semantic audio transformer for sound classification and detection.
\newblock In \emph{IEEE International Conference on Acoustics, Speech and Signal Processing, ICASSP}, 2022.

\bibitem[Dong et~al.(2023)Dong, Takahashi, Mitsufuji, McAuley, and Berg-Kirkpatrick]{dong2023clipsep}
Hao-Wen Dong, Naoya Takahashi, Yuki Mitsufuji, Julian McAuley, and Taylor Berg-Kirkpatrick.
\newblock Clipsep: Learning text-queried sound separation with noisy unlabeled videos.
\newblock In \emph{Proceedings of International Conference on Learning Representations (ICLR)}, 2023.

\bibitem[Gao and Grauman(2019)]{gao2019co}
Ruohan Gao and Kristen Grauman.
\newblock Co-separating sounds of visual objects.
\newblock In \emph{Proceedings of the IEEE/CVF International Conference on Computer Vision}, pages 3879--3888, 2019.

\bibitem[Guo et~al.(2024)Guo, Qu, Niu, Qi, Yue, Shi, Xing, and Ying]{guo2024open}
Ruohao Guo, Liao Qu, Dantong Niu, Yanyu Qi, Wenzhen Yue, Ji~Shi, Bowei Xing, and Xianghua Ying.
\newblock Open-vocabulary audio-visual semantic segmentation.
\newblock In \emph{Proceedings of the 32nd ACM International Conference on Multimedia}, pages 7533--7541, 2024.

\bibitem[Huang et~al.(2012)Huang, Chen, Smaragdis, and Hasegawa-Johnson]{huang2012singing}
Po-Sen Huang, Scott~Deeann Chen, Paris Smaragdis, and Mark Hasegawa-Johnson.
\newblock Singing-voice separation from monaural recordings using robust principal component analysis.
\newblock In \emph{2012 IEEE International Conference on Acoustics, Speech and Signal Processing (ICASSP)}, pages 57--60. IEEE, 2012.

\bibitem[Islam et~al.(2024)Islam, Nabavi, Kezele, Wang, Yu, and Tang]{Islam2024VGASS}
Md.~Amirul Islam, Seyed~Shahabeddin Nabavi, Irina Kezele, Yang Wang, Yuanhao Yu, and Jin Tang.
\newblock Visually guided audio source separation with meta consistency learning.
\newblock In \emph{Proceedings of the IEEE/CVF Winter Conference on Applications of Computer Vision (WACV)}, pages 3014--3023, 2024.
\newblock URL \url{https://openaccess.thecvf.com/content/WACV2024/papers/Islam_Visually_Guided_Audio_Source_Separation_With_Meta_Consistency_Learning_WACV_2024_paper.pdf}.

\bibitem[Kong et~al.(2021)Kong, Cao, Liu, Choi, and Wang]{kong2021decoupling}
Qiuqiang Kong, Yin Cao, Haohe Liu, Keunwoo Choi, and Yuxuan Wang.
\newblock Decoupling magnitude and phase estimation with deep resunet for music source separation.
\newblock \emph{arXiv preprint arXiv:2109.05418}, 2021.

\bibitem[Liang et~al.(2022)Liang, Zhang, Kwon, Yeung, and Zou]{ModalityGap}
Weixin Liang, Yuhui Zhang, Yongchan Kwon, Serena Yeung, and James Zou.
\newblock Mind the gap: Understanding the modality gap in multi-modal contrastive representation learning.
\newblock In \emph{NeurIPS}, 2022.
\newblock URL \url{https://openreview.net/forum?id=S7Evzt9uit3}.

\bibitem[Mo and Tian(2024)]{mo2024semanticgroupingnetworkaudio}
Shentong Mo and Yapeng Tian.
\newblock Semantic grouping network for audio source separation, 2024.
\newblock URL \url{https://arxiv.org/abs/2407.03736}.

\bibitem[Molla and Hirose(2007)]{molla2007single}
Md~Khademul~Islam Molla and Keikichi Hirose.
\newblock Single-mixture audio source separation by subspace decomposition of hilbert spectrum.
\newblock \emph{IEEE Transactions on Audio, Speech, and Language Processing}, 15\penalty0 (3):\penalty0 893--900, 2007.

\bibitem[Nakada et~al.(2024)Nakada, Nishimura, Munakata, Kondo, and Komatsu]{nakada2024deteclap}
Shota Nakada, Taichi Nishimura, Hokuto Munakata, Masayoshi Kondo, and Tatsuya Komatsu.
\newblock Deteclap: Enhancing audio-visual representation learning with object information.
\newblock \emph{arXiv preprint arXiv:2409.11729}, 2024.

\bibitem[Nath and Sarma(2024)]{nath2024separation}
Kakali Nath and Kandarpa~Kumar Sarma.
\newblock Separation of overlapping audio signals: a review on current trends and evolving approaches.
\newblock \emph{Signal Processing}, page 109487, 2024.

\bibitem[Petermann et~al.(2020)Petermann, Chandna, Cuesta, Bonada, and Gomez]{petermann2020deeplearningbasedsource}
Darius Petermann, Pritish Chandna, Helena Cuesta, Jordi Bonada, and Emilia Gomez.
\newblock Deep learning based source separation applied to choir ensembles, 2020.
\newblock URL \url{https://arxiv.org/abs/2008.07645}.

\bibitem[Radford et~al.(2021)Radford, Kim, Hallacy, Ramesh, Goh, Agarwal, Sastry, Askell, Mishkin, Clark, Krueger, and Sutskever]{radford2021learningtransferablevisualmodels}
Alec Radford, Jong~Wook Kim, Chris Hallacy, Aditya Ramesh, Gabriel Goh, Sandhini Agarwal, Girish Sastry, Amanda Askell, Pamela Mishkin, Jack Clark, Gretchen Krueger, and Ilya Sutskever.
\newblock Learning transferable visual models from natural language supervision, 2021.
\newblock URL \url{https://arxiv.org/abs/2103.00020}.

\bibitem[Ronneberger et~al.(2015)Ronneberger, Fischer, and Brox]{ronneberger2015u}
Olaf Ronneberger, Philipp Fischer, and Thomas Brox.
\newblock U-net: Convolutional networks for biomedical image segmentation.
\newblock In \emph{Medical image computing and computer-assisted intervention--MICCAI 2015: 18th international conference, Munich, Germany, October 5-9, 2015, proceedings, part III 18}, pages 234--241. Springer, 2015.

\bibitem[Stoller et~al.(2018)Stoller, Ewert, and Dixon]{stoller2018wave}
Daniel Stoller, Sebastian Ewert, and Simon Dixon.
\newblock Wave-u-net: A multi-scale neural network for end-to-end audio source separation.
\newblock \emph{arXiv preprint arXiv:1806.03185}, 2018.

\bibitem[Tian et~al.(2021)Tian, Hu, and Xu]{tian2021cyclic}
Yapeng Tian, Di~Hu, and Chenliang Xu.
\newblock Cyclic co-learning of sounding object visual grounding and sound separation.
\newblock In \emph{Proceedings of the IEEE/CVF Conference on Computer Vision and Pattern Recognition (CVPR)}, pages 2745--2754, 2021.

\bibitem[Vincent et~al.(2006)Vincent, Gribonval, and Fevotte]{metrics06}
E.~Vincent, R.~Gribonval, and C.~Fevotte.
\newblock Performance measurement in blind audio source separation.
\newblock \emph{IEEE Transactions on Audio, Speech, and Language Processing}, 14\penalty0 (4):\penalty0 1462--1469, 2006.
\newblock \doi{10.1109/TSA.2005.858005}.

\bibitem[Virtanen(2007)]{virtanen2007monaural}
Tuomas Virtanen.
\newblock Monaural sound source separation by nonnegative matrix factorization with temporal continuity and sparseness criteria.
\newblock \emph{IEEE transactions on audio, speech, and language processing}, 15\penalty0 (3):\penalty0 1066--1074, 2007.

\bibitem[Wu et~al.(2022)Wu, Seetharaman, Kumar, and Bello]{wu2022wav2clip}
Ho-Hsiang Wu, Prem Seetharaman, Kundan Kumar, and Juan~Pablo Bello.
\newblock Wav2clip: Learning robust audio representations from clip.
\newblock In \emph{ICASSP 2022 - 2022 IEEE International Conference on Acoustics, Speech and Signal Processing (ICASSP)}, 2022.

\bibitem[Wu* et~al.(2023)Wu*, Chen*, Zhang*, Hui*, Berg-Kirkpatrick, and Dubnov]{laionclap2023}
Yusong Wu*, Ke~Chen*, Tianyu Zhang*, Yuchen Hui*, Taylor Berg-Kirkpatrick, and Shlomo Dubnov.
\newblock Large-scale contrastive language-audio pretraining with feature fusion and keyword-to-caption augmentation.
\newblock In \emph{IEEE International Conference on Acoustics, Speech and Signal Processing, ICASSP}, 2023.

\bibitem[Yu et~al.(2019)Yu, Lin, Yang, Shen, Lu, and Huang]{yu2019free}
Jiahui Yu, Zhe Lin, Jimei Yang, Xiaohui Shen, Xin Lu, and Thomas~S Huang.
\newblock Free-form image inpainting with gated convolution.
\newblock In \emph{Proceedings of the IEEE/CVF international conference on computer vision}, pages 4471--4480, 2019.

\bibitem[Zhao et~al.(2018)Zhao, Gan, Rouditchenko, Vondrick, McDermott, and Torralba]{Zhao_2018_ECCV}
Hang Zhao, Chuang Gan, Andrew Rouditchenko, Carl Vondrick, Josh McDermott, and Antonio Torralba.
\newblock The sound of pixels.
\newblock In \emph{The European Conference on Computer Vision (ECCV)}, September 2018.

\bibitem[Zhao et~al.(2019)Zhao, Gan, Ma, and Torralba]{zhao2019sound}
Hang Zhao, Chuang Gan, Wei-Chiu Ma, and Antonio Torralba.
\newblock The sound of motions.
\newblock In \emph{Proceedings of the IEEE/CVF International Conference on Computer Vision}, pages 1735--1744, 2019.

\end{thebibliography}

\vspace{-2mm}
\clearpage

\setcounter{section}{0}
\setcounter{table}{0}
\setcounter{figure}{0}
\setcounter{page}{1}

\setcounter{section}{0}
\setcounter{table}{0}
\setcounter{figure}{0}
\setcounter{page}{1}

~
\vspace{-10mm}
\section*{\huge Appendix}

\renewcommand{\thefigure}{A\arabic{figure}}
\renewcommand{\thetable}{A\arabic{table}}
\renewcommand{\thesection}{A\arabic{section}}

\renewcommand{\theHsection}{section.S\arabic{section}}
\renewcommand{\theHfigure}{figure.S\arabic{figure}}
\renewcommand{\theHtable}{table.S\arabic{table}}

\section{Acoustic Characteristics Computation for Figure 1}
\label{app:s1}

In Figure \ref{fig:instrument_fusion}, we define two acoustic dimensions for each instrument class: Transient Property and Harmonic Complexity based on two musically meaningful metrics.
These metrics are inspired by concepts in music theory and psychoacoustics \cite{caclin2005acoustic}: 
transient property reflects attack sharpness or percussiveness, while harmonic complexity relates to timbral richness and overtone distribution.

\paragraph{Transient Property.} We define the \textit{Amplitude Ratio (AR)} for a waveform $x(t)$ as:
\[
\text{AR} = \frac{\max(|x(t)|)}{\sqrt{\frac{1}{T} \sum_{t=1}^{T} x(t)^2}}
\]
A higher AR indicates a more transient, percussive sound. We compute AR for all samples in each instrument class for MUSIC dataset and take the average to represent the class.

\paragraph{Harmonic Complexity.} Given an audio clip, we estimate its fundamental frequency $f_0$ using \texttt{librosa.pyin}, and compute average energy $E_i$ at its first $N=8$ harmonics. The Harmonic Component Ratio (HCR) is then defined as:
\[
\text{HCR} = \frac{\sum_{i=1}^{N} \frac{1}{i} \cdot E_i}{\sum_{i=1}^{N} E_i}
\]
We define harmonic complexity as $1 - \text{HCR}$, so that larger values represent richer high-order harmonic content. Final y-axis values in Figure 1 are computed by averaging this score across all clips of each class in the MUSIC dataset.

\section{More Audio Separation Examples}
\label{app:more_examples}

\begin{figure}
\centering
\begin{tabular}{c}
\bmvaHangBox{%
    \includegraphics[width=0.95\linewidth]{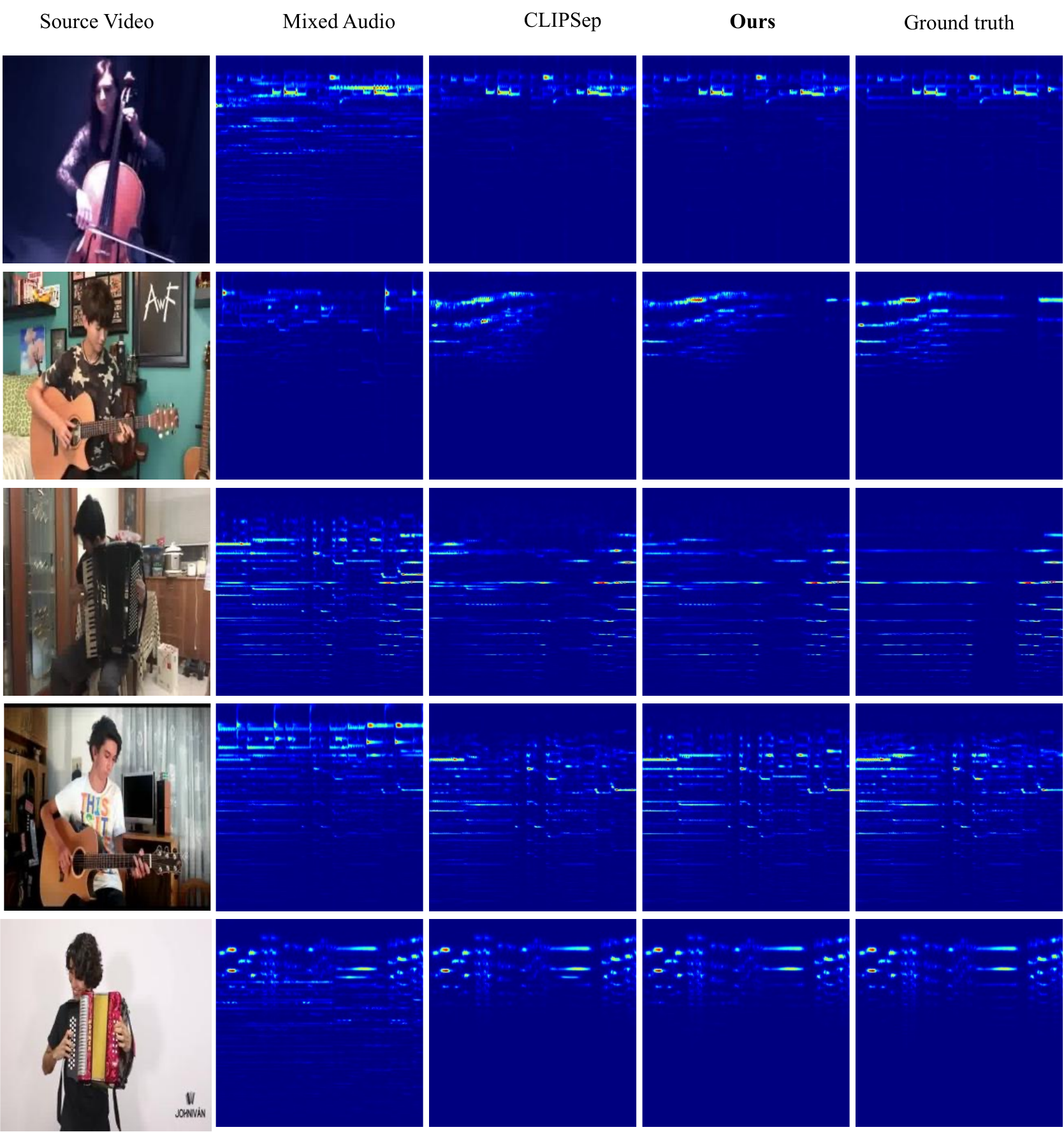}%
}\\
\end{tabular}
\caption{\textbf{More Visualisation Examples on MUSIC dataset.}
We visualise the spectrograms for original audio, mixed audio, and predictions from different audio separation models with corresponding source query frames. We compared our method (the fourth column) with Clipsep (the third column).}
\label{fig:s1}
\end{figure}

Qualitative results in Figure~\ref{fig:s1} show that our method generates spectrograms with fewer artefacts and better alignment with the ground truth.

\section{A Comparative Analysis of Fusion Stage for Audio-Visual Separation}
\label{sec:s3}

\begin{table}[htbp]
\centering
\caption{\textbf{Instrument-wise separation performance (SDR) comparison of three fusion strategies on the MUSIC dataset.} 11 solo instruments are analysed. For fair comparison, no representation alignment is used here. Best results in \textbf{bold}.}
\label{tab:instrument_results}
\begin{tabular}{lccc}
\toprule
\multirow{2}{*}{\textbf{Instrument Type}} & \multicolumn{3}{c}{\textbf{Fusion Method}} \\
\cmidrule(lr){2-4}
& \textbf{Middle} & \textbf{Late} & \textbf{Hierarchical} \\
\midrule
Violin          & \textbf{6.49} & 6.16 & 4.58 \\
Clarinet        & 3.93 & 4.13 & \textbf{7.56} \\
Saxophone       & 4.11 & 5.79 & \textbf{8.94} \\
Acoustic Guitar & 6.38 & \textbf{8.26} & 7.57 \\
Xylophone       & 8.14 & \textbf{13.49} & 11.43 \\
Flute           & 2.01 & 1.57 & \textbf{4.96} \\
Accordion       & 4.51 & 5.65 & \textbf{5.84} \\
Cello           & 4.21 & 4.85 & \textbf{5.19} \\
Erhu            & -0.07 & \textbf{1.95} & 0.61 \\
Tuba            & 6.25 & 3.49 & \textbf{7.73} \\
Trumpet         & \textbf{13.87} & 12.54 & 13.62 \\
\midrule
\textbf{Overall SDR} & 5.57 & 5.86 & \textbf{6.65} \\
\bottomrule
\end{tabular}
\end{table}

Table~\ref{tab:instrument_results} presents the SDR results for different fusion strategies across various musical instruments. 
We find that middle fusion is particularly effective for instruments with transient sounds or simple spectral structures. 
Specifically, instruments such as the trumpet and flute produce short-duration notes, where the energy is concentrated in localised time windows in the time-frequency representation. When visual information is fused at the bottleneck of the U-Net, it deeply influences the entire decoding process and provides benefits for enhancing localised features, which in turn improves separation performance for short transient sounds.  

Complex harmonic structures are common in instruments such as acoustic guitars, xylophones, and accordions, where the sound is composed of multiple overtones. The mathematical formulation of a harmonic-rich signal is:
\[
    X(\omega) = \sum_{n=1}^{N} A_n e^{j\omega t_n},
\]
where each harmonic component \( A_n e^{j\omega t_n} \) contributes to the overall timbre of the sound.

This may partially explain why incorporating visual cues at the bottleneck of the audio U-Net can interfere with the decoder's ability to capture intricate harmonic relationships, potentially leading to the loss or distortion of certain harmonics.

Late fusion is more effective for separating instruments that produce sustained tones or have complex harmonic structures. 
Unlike middle fusion, which influences the separation process from an early stage, late fusion applies visual weighting only at the final stage of the decoder. Since it does not interfere with most of the decoding process, the model can reconstruct the audio with the information only from the audio modality before making fine-grained adjustments using visual information at the output. This ensures that the model preserves the temporal coherence of long-duration notes, preventing distortions that could disrupt the smooth progression of sustained sounds.

\end{document}